# Insight Rumors: A Novel Textual Rumor Locating and Marking Model Leveraging Att_BiMamba2 Network

## Authors

Bin Ma, Yifei Zhang, Yongjin Xian, Qi Li, Linna Zhou, Gongxun Miao

## Abstract

With the development of social media networks, rumor detection models have attracted more and more attention. Whereas, these models primarily focus on classifying contexts as rumors or not, lacking the capability to locate and mark specific rumor content. To address this limitation, this paper proposes a novel rumor detection model named Insight Rumors to locate and mark rumor content within textual data. Specifically, we propose the Bidirectional Mamba2 Network with Dot-Product Attention (Att_BiMamba2), a network that constructs a bidirectional Mamba2 model and applies dot-product attention to weight and combine the outputs from both directions, thereby enhancing the representation of high-dimensional rumor features. Simultaneously, a Rumor Locating and Marking module is designed to locate and mark rumors. The module constructs a skip-connection network to project high-dimensional rumor features onto low-dimensional label features. Moreover, Conditional Random Fields (CRF) is employed to impose strong constraints on the output label features, ensuring accurate rumor content location. Additionally, a labeled dataset for rumor locating and marking is constructed, with the effectiveness of the proposed model is evaluated through comprehensive experiments. Extensive experiments indicate that the proposed scheme not only detects rumors accurately but also locates and marks them in context precisely, outperforming state-of-the-art schemes that can only discriminate rumors roughly.

## 1. Introduction

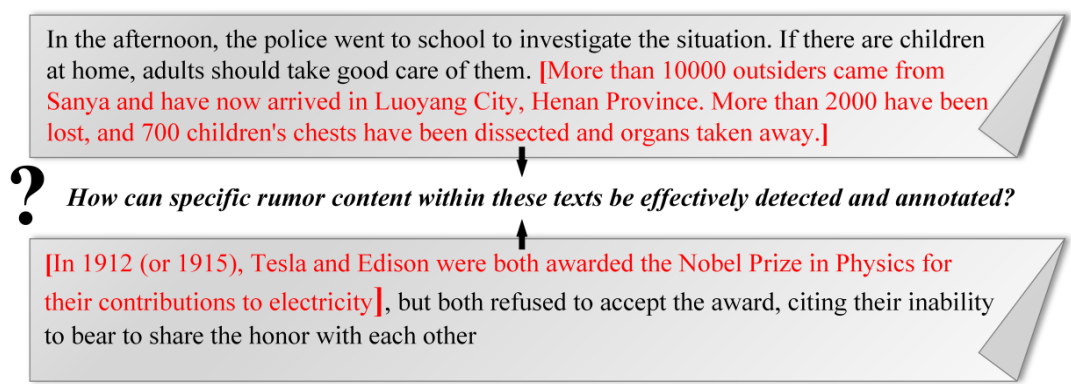

Figure 1. The problem that "Insight Rumors" aims to solve.

In the field of rumor detection research, various methods have made substantial progress in classifying rumors. FakeKG [Shahi and Kishore, 2023] introduced a system that enhances automatic fact-checking by using knowledge graphs. By building a knowledge graph of false statements, it boosts the efficiency and accuracy of large-scale fact verification. Another research [Si *et al.*, 2022] explored faithful reasoning for multi-hop fact verification through the use of salience-aware graph learning techniques. In addition, HG-SL [Sun *et al.*, 2023] proposed a model for early fake news detection by jointly learning global and local user propagation behaviors. It effectively integrates both global user behavior information and local details, enhancing detection performance. Event-Radar [Ma *et al.*, 2024] uses multi-view learning for multimodal fake news detection. It introduces an event-driven learning framework that further enhances the processing and integration of multimodal data. Other studies, such as the evidence retrieval method presented by [Zheng *et al.*, 2024], also demonstrate the importance of evidence in fact verification, stressing that retrieving relevant evidence can almost entirely resolve the issue of fact-checking. Moreover, evidence-enhanced reasoning frameworks [Wu *et al.*,2024] and natural language-based reasoning networks [Zhang *et al.*, 2024] have further advanced the development of fake news detection technologies, particularly in the application of multimodal fake news detection. [Liu *et al.*, 2024] examined the transition from skepticism to acceptance by simulating the dynamics of attitudes during the propagation of fake news, shedding light on the complexity of the mechanisms of fake news spread. Although these models can effectively determine whether the content of the data is a rumor, they generally lack in-depth detection and detailed locating and marking of specific rumor content. This means that existing systems are often able to determine whether a piece of information is a rumor, but they struggle to further analyze its specific false content and influencing factors. This limitation restricts the depth and scope of rumor analysis, impeding the formulation and implementation of targeted counter-strategies.

The locating and marking of specific rumor content are of significant importance. First, meticulous content analysis allows researchers to gain deeper insights into the true nature of rumors. Second, detailed locating and marking provide fact-checkers with specific references for verifying information, thereby improving verification efficiency. Moreover, it contributes to the establishment of a comprehensive rumor monitoring network, thereby safeguarding information security. In addition, in-depth content detection provides trustworthy information, thus enhancing society's ability to recognize and resist rumors. Therefore, conducting locating and marking of specific rumor content is not only a key step in advancing rumor detection technology but also a crucial measure for ensuring the safety of information dissemination.

At present, sequence labeling models are advancing rapidly, and many innovative methods are driving progress in this area. For example, [Yan *et al.*, 2023] proposed modeling nested named entity recognition as a local hypergraph structure, which further enhances the ability to recognize complex nested structures. Another research [Cui and Zhang, 2019] proposed the Hierarchically-Refined Label Attention Network, which effectively boosts the performance of sequence labeling tasks, particularly in processing long sequences and complex label relationships. [Wang *et al.*, 2020] advanced the cross-lingual transfer ability of multilingual sequence labeling models using the Structure-Level Knowledge Distillation method. In addition, the Bi-directional LSTM-CNN-CRF model proposed by [Ma and Hovy, 2016] offers an end-to-end solution for sequence labeling, which is extensively applied in named entity recognition and other sequence labeling tasks. The effective method for Chinese named entity recognition proposed by [Gu *et al.*, 2022] further improves the precision and robustness of Chinese NER by deeply mining the regularities of Chinese corpora. Other research, like the fine-grained knowledge fusion method presented by [Yang *et al.*, 2019], offers effective solutions to domain adaptation issues in the sequence labeling field. Despite the improvements in locating and marking accuracy brought by these methods, sequence labeling tasks still encounter problems like handling long-distance dependencies, key information loss, data scarcity, poor locating and marking quality, and low computational efficiency.

Facing the lack of detailed locating and marking of specific rumor content in rumor detection research, along with challenges in sequence labeling tasks, such as long-range dependencies, key information loss, data scarcity, poor locating and marking quality, and low computational efficiency, this paper treats the detection and locating and marking of rumor content as a specialized sequence labeling problem and proposes the "Insight Rumors" model to locate and mark specific rumor content in text, addressing the issues outlined in Figure 1. To the best of our knowledge, this is the first model focused on the detection and detailed locating and marking of specific rumor content. The model first uses a pre-trained BERT model to encode the text into word vector sequences. It then constructs a bidirectional Mamba2 model and applies dot-product attention to weigh and combine the outputs from both directions, obtaining a rumor feature vector. Next, the paper designs a skip-connection network to project high-dimensional rumor features onto low-dimensional label features, minimizing the loss of rumor information during the dimensionality reduction process. Finally, the model employs Conditional Random Fields (CRF) to impose strong constraints on the label features, achieving more accurate rumor content locating and marking. Through comparative experiments and ablation tests, we validate the model's efficiency and the effectiveness of its individual components.

The main contributions of this paper are as follows:

●In contrast to existing rumor detection algorithms that only achieve rough classification, we introduce the "Insight Rumors" model, which is based on the Bidirectional Mamba2 Network with Dot-Product Attention (Att_BiMamba2) to precisely locate and mark specific rumor content through in-depth analysis of text sequences.

●A bidirectional Mamba2 model with attention, Att_BiMamba2, is proposed, which simultaneously learns rumor features from both the forward and backward directions of the sequence. It also employs dot-product attention to assess the importance of outputs from both directions and performs weighted summation to enhance the expressive power of the output features for rumor information.

●A Rumor Locating and Marking module is designed for rumor locating and marking. This module first constructs a skip-connection network to project high-dimensional rumor features onto low-dimensional label features, minimizing information loss during the projection process. It then incorporates CRF to impose strong constraints on the low-dimensional label features, enhancing the accuracy of rumor locating and marking.

●A new dataset, IR-WEIBO, has been constructed for locating and marking specific rumor content. We improved existing sequence labeling methods, enabling them to perform this task, with compared their performance in locating and marking rumors with the proposed model. The results validate the effectiveness and superiority of the proposed model.

# 2. Related Work

## 2.1. Bidirectional Encoder Representations from Transformers

BERT (Bidirectional Encoder Representations from Transformers) has made significant progress in the field of Natural Language Processing (NLP) by introducing the bidirectional encoder representation model. It was proposed by Devlin et al. (2019) and has become the cornerstone of many NLP tasks, including Named Entity Recognition (NER), sentiment analysis, and text classification. BERT adopts the Transformer architecture, which enables bidirectional context modeling, making it more powerful than traditional unidirectional language models. In addition, it has been widely studied for its application in sequence labeling tasks such as Named Entity Recognition and Part-of-Speech tagging [Devlin *et al.*, 2019]. The "Bidirectional" in BERT refers to its pretraining process, where the bidirectional Transformer allows each word's representation to be influenced by both the preceding and succeeding words. This approach captures more contextual information, making the model more accurate in tasks such as word sense disambiguation and Named Entity Recognition.

## 2.2. Mamba2

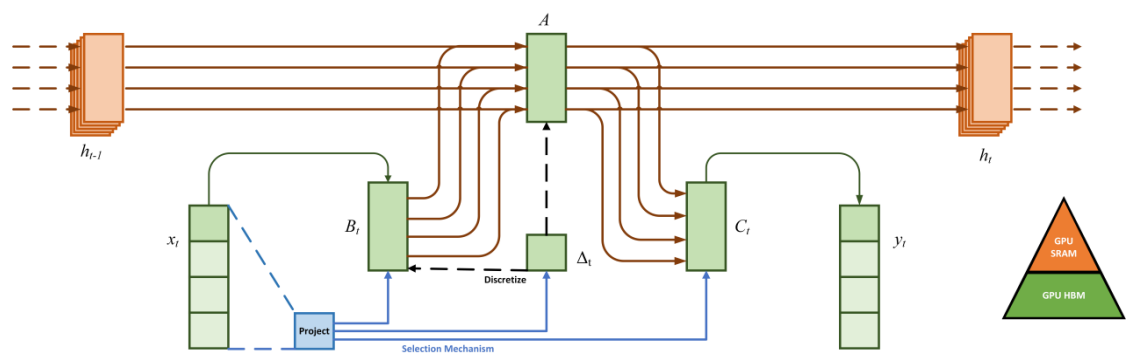

Figure 2.The Mamba model architecture.

Mamba is a novel model based on State Space Models (SSMs), which achieves efficient processing of long sequence data by introducing a selective state space mechanism. As shown in Figure 2, the core of the model is its ability to dynamically adjust its parameters based on the input data, enabling selective attention to or ignoring of specific information. This capability allows it to excel in processing complex data like language, audio, and genomics. Another distinctive feature of Mamba is its hardware-aware parallel algorithm, which optimizes the utilization of GPU memory hierarchy, significantly enhancing the computational efficiency of the model. Additionally, Mamba adopts a simplified architecture design that integrates the previously separate structured state space model and multilayer perceptron blocks, further enhancing the model's performance and flexibility [Albert and Dao, 2023]. While maintaining linear time complexity, Mamba can achieve or surpass the performance of existing Transformer models in various tasks, especially when handling ultra-long sequences, where its advantages are even more pronounced.

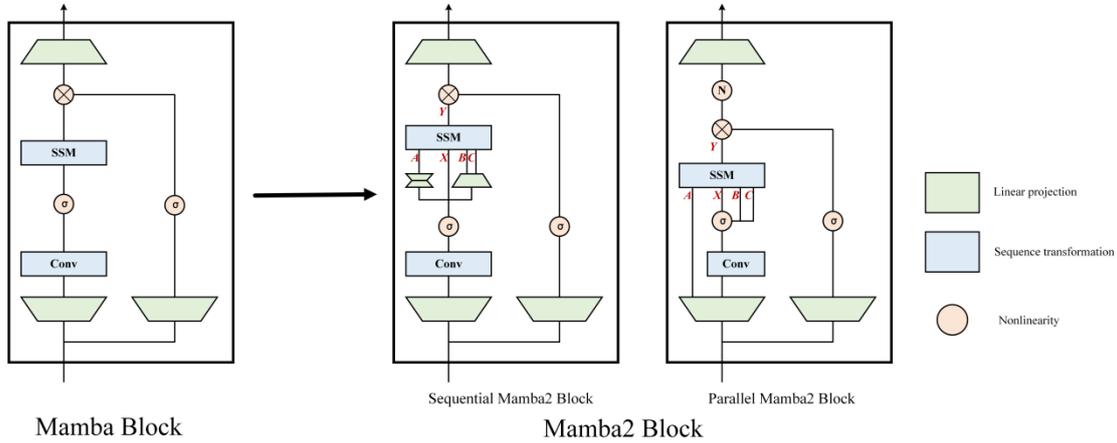

Mamba Block  Mamba2 Block

Figure 3. The internal structures of Mamba and Mamba2 modules.

Mamba2, proposed based on Mamba [Dao and Gu, 2024], aims to enhance sequence modeling efficiency and performance by introducing a theoretical connection between structured state space models (SSM) and attention mechanisms. As shown in Figure 3, Mamba2 introduces context-aware mechanisms and multi-layer attention mechanisms. It allows dynamic adjustment of state transition parameters based on the input, enabling selective processing of information. Compared to Mamba, Mamba2 simplifies the model structure by using parallel parameter projection and additional normalization layers, reducing instability during the training process and improving computational efficiency. Moreover, Mamba2 adopts a new hardware-friendly algorithm that utilizes the characteristics of structured matrices, enabling more efficient matrix multiplication calculations on modern hardware.

## 2.3. Conditional Random Field

Conditional Random Field (CRF), introduced by [Lafferty *et al.*, 2001], is used for labeling and segmenting sequence data and has found widespread application in tasks like named entity recognition, part-of-speech tagging, and chunking. The key advantage of the CRF model is its discriminative property, which directly models the conditional probability of the label sequence. It describes the relationship between the input and the labels by defining a set of feature functions, which can be either transition features or emission features, and these together form the model's observation function. During the model training phase, CRF learns parameters through maximum likelihood estimation to maximize the log-likelihood function of the training data. In the prediction phase, CRF uses the learned model parameters and dynamic programming algorithms, such as the Viterbi algorithm, to find the label sequence with the highest conditional probability for a given input sequence. Due to its flexibility and effectiveness, CRF has been widely applied in natural language processing, particularly in tasks like part-of-speech tagging and named entity recognition. However, CRF typically relies on manually designed features and has certain limitations in capturing long-distance dependencies. Therefore, many studies combine CRF with deep learning architectures (such as LSTM and CNN) to propose more powerful hybrid models. For instance, [Lample *et al.*, 2016] combined CRF with bidirectional LSTM to propose a model for named entity recognition, achieving outstanding performance.

# 3. Problem Definition

Rumor detection is typically defined as a binary classification problem to determine whether the content of a text description is a rumor. However, for the task of detecting and labeling specific rumor content within the text, it is necessary to label the specific rumor content within it. Therefore, we define this problem as a specialized sequence labeling process, where each element in the sequence is labeled, i.e., performing multi-class classification for each element in the sequence.

Consider the input is a text sequence $X = \{x_1, x_2, \ldots, x_n\}$, consisting of $n$ elements, where each element $x_i$ could be a word, character, timestamp, etc. In this task, we aim to predict for each element $x_i$ in the text sequence whether it belongs to the rumor content and assign a corresponding label $y_i$. Thus, the output is a label sequence $Y = \{y_1, y_2, \ldots, y_n\}$, where each label $y_i$ indicates whether $x_i$ is part of the rumor.

To achieve this task, we define the labels as follows:
$$Label = \{B - Rumor, I - Rumor, O\} \tag{1}$$
Where *B-Rumor* indicates the beginning of rumor content, *I-Rumor* indicates the rumor content, and *O* indicates non-rumor content. The *B-Rumor* tag serves to effectively delineates the boundaries of rumors. For instance, when two adjacent sentences in a description both contain rumor content—where the end of the first sentence and the beginning of the second sentence both contain rumor content—they should be labeled as two separate rumor entities, not as a single whole. In the experiments, we map the labels as follows: *B-Rumor* to 0, *I-Rumor* to 1, and *O* to 2.

Specifically, the objective of this task is to learn a mapping function $F$ such that, given an input text sequence $X$, it can accurately predict the corresponding label sequence $Y$, i.e:
$$F: X = \{x_1, x_2, \ldots, x_n\} \rightarrow Y = \{y_1, y_2, \ldots, y_n\}, y_i \in Label \tag{2}$$
Where $F$ is a method capable of making predictions based on the contextual information of the input sequence (such as surrounding words, features, etc.).

# 4. Methodology

Figure 4 provides an overview of the solution to the problem of locating and marking rumor content in text descriptions. In this chapter, Section 4.1 outlines the overall framework of the model; Section 4.2 describes how to obtain the word vectors for each element in the text sequence; Section 4.3 introduces the improved Mamba2 model, Bidirectional Mamba2 Network with Dot-Product Attention (Att_BiMamba2), which learns rumor features from the word vectors.; Section 4.4 provides a detailed explanation of the Rumor Locating and Marking module, where we construct a Skip-connection network to ensure the integrity of rumor information during the mapping from high-dimensional rumor features to low-dimensional label features, and apply CRF to impose strong constraints for more precise locating and marking; Section 4.5 discusses the loss function employed for model optimization.

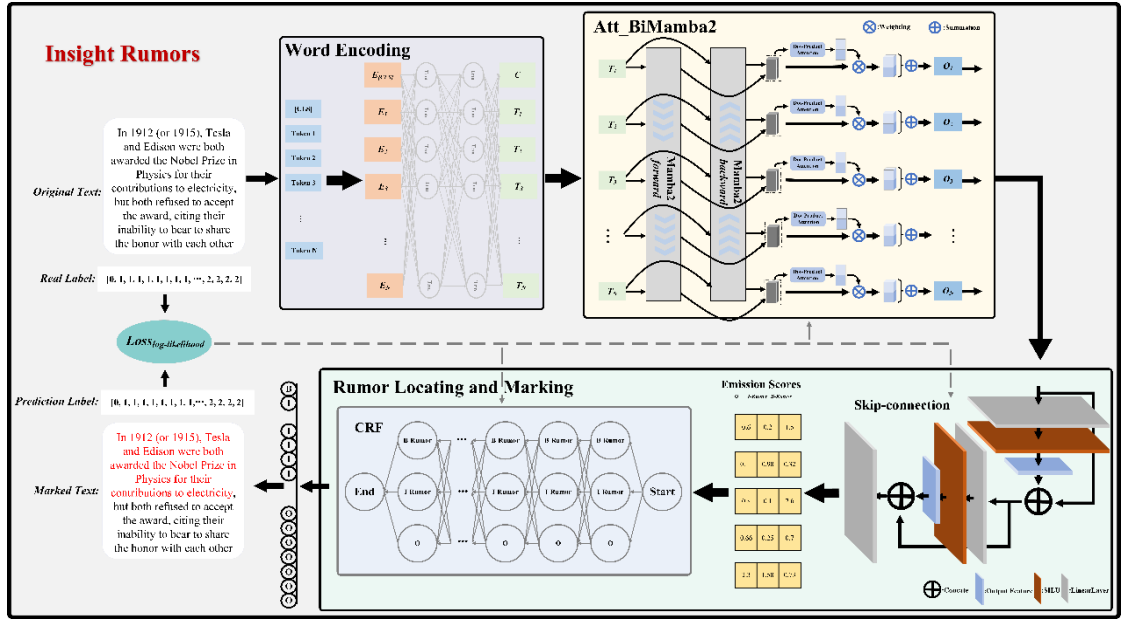

Figure 4. The framework structure of Insight Rumors, primarily including: word sequence encoding, rumor feature extraction using the Att_BiMamba2 network, Rumor Locating and Marking.

## 4.1. Framework

The framework of Insight Rumors is illustrated in Figure 4, consisting of three main parts: Word Encoding, Bidirectional Mamba2 Network with Dot-Product Attention (Att_BiMamba2), and Rumor Locating and Marking . The model first divides the text into a token sequence and employs the pre-trained BERT for encoding to obtain the word vectors $T_i$ for each token. Then, Insight Rumors constructs a bidirectional Mamba2 network to learn contextual rumor features from both directions of the sequence, and applies dot-product attention to assess the importance of the outputs of Mamba2 in different directions. The evaluated scores are then used to weight and sum the outputs, yielding stronger feature representations ($O_i$) for rumor information. To ensure the integrity of the mapping from high-dimensional rumor features to low-dimensional label features, Insight Rumors employs skip-connection in the Rumor Locating and Marking module to gradually reduce dimensions, minimizing the loss of rumor information during the mapping process. Finally, Conditional Random Fields (CRF) are applied to impose strong constraints on labeling process by learning the parameter transition matrix, effectively improving the accuracy of sequence labeling.

## 4.2. Word Encoding

Given a text sequence $X = \{x_1, x_2, ..., x_n\}$, where $x_i$ represents the i-th token, the first step is to obtain the Token Embedding, i.e., find the embedding representation $TE_{x_i}$ for each token $x_i$. If the text contains multiple sentences, a segment embedding $S_{x_i}$ is assigned to each sentence to identify which sentence each word belongs to, marked as Segment Embedding. To preserve the order information, a position embedding $P_{x_i}$ is provided for each token during the encoding process, indicating the position of token $x_i$ in the sequence, marked as Positional Embedding. After combining these three embeddings, the initial input representation for each token is obtained:

$$E = [TE_{x_1} + S_{x_1} + P_{x_1}, TE_{x_2} + S_{x_2} + P_{x_2}, \ldots, TE_{x_n} + S_{x_n} + P_{x_n}] \quad (3)$$

These initial representations are input into a 12-layer Transformer encoder. In each layer, the self-attention mechanism captures the relationships between tokens. Specifically, for the input $H_{l-1}$ of the $l$-th layer, the self-attention mechanism calculates the new representation of each word based on queries (Query), keys (Key), and values (Value). The calculation formula for self-attention is as follows:

$$Attention(Q, K, V) = softmax\left(\frac{QK^T}{\sqrt{d_k}}\right)V \quad (4)$$

In addition, after each self-attention module, a feedforward neural network further processes the representation of each word. The calculation method of the feedforward network is as follows:

$$Feed\ Forward(h) = max(0, W_1 h + b_1) W_2 + b_2 \quad (5)$$

After the computation through multiple layers of the Transformer encoder, the output of the last layer will contain the context-dependent representation of each word. This output represents the semantic information of each word in its context, capturing the polysemy of words and the complex relationships between words. The word representation extracted from $H_L$, denoted as $T_i$, is the context-dependent feature of each word in the text sequence.

$$T = \{T_1, T_2, \ldots, T_N\} \quad (6)$$

Where $N$ is the length of the text sequence.

## 4.3. Bidirectional Mamba2 Network with Dot-Product Attention

The Bidirectional Mamba2 Network with Dot-Product Attention (Att_BiMamba2) is constructed to learn the rumor features from the word features $T$ in the text. The architecture of this network is shown in Figure 4. The input to the network is the sequence of word features $T$ from the text. The network is composed of two key components: Bi_Mamba2 and Dot-Product Attention.

**BiMamba2:** The architecture of the Mamba2 Block is shown in Figure 3. Building on this architecture, we construct BiMamba2, which models the long-range dependencies of each word feature using bidirectional SSM, capturing rumor information more comprehensively from both directions of the sequence.

Initially, the original input $T$ is adjusted through a fully connected layer to align with the internal representation dimensions of the model. The adjusted input $x_{adjusted}$ is then passed into the forward Mamba2 layer (Mamba2$_{forward}$).

$$x_{adjusted} = fc\_in(T) \quad (7)$$
$$x_{forward} = Mamba2_{forward}(x_{adjusted}) \quad (8)$$

By flipping along the time-step dimension of the sequence, the adjusted sequence is fed into the reverse Mamba2 layer. This step simulates the sequence processing of the reverse Mamba2, enabling the model to comprehend the sequence data from both directions, thus enhancing its capability to handle long-range dependencies.

$$x_{backward} = Mamba2_{backward}\left(flip(x_{adjusted}, time)\right) \quad (9)$$

Here, $flip(x_{adjusted}, time)$ refers to the reversal of the sequence $x_{adjusted}$ along the time dimension

Specifically, the input is projected first, and the feature mapping of the input is calculated to capture more intricate and complex features.

$$Z = Linear(X) \quad (10)$$

The dimension of Z is $N \times D_{inner}$, where $D_{inner}$ is the expanded dimension.

The deep separable convolution is used to process features along the time dimension with a 1D convolution operation, aimed at extracting local features and enhancing the local representation power of the features. Assuming the kernel size is $k$, the coverage range defines the local temporal dependency area captured by the module. The result is then processed through the SILU activation function. The convolution process can be represented as:

$$X_{conv} = Conv1D(Z, kernel\_size = k) \tag{11}$$

$$X_{conv\_activated} = SILU(X_{conv}) \tag{12}$$

The State Space Model (SSM) further models the input feature $X_{conv\_activated}$, capturing the dependencies along the time dimension. Through the sparsification or low-rank constraints on matrices $A$, $B$, and $C$, SSM can capture long-range temporal dependencies with low computational cost, making it suitable for sequence data modeling.

$$Y = SSM(A, B, C; X_{conv\_activated}) \tag{13}$$

In this process, $A$ captures the global dependencies between time steps, akin to the $QK^T$ calculation in attention mechanisms. The input $x_t$ is mapped to the hidden state space, allowing it to participate in the state update. Similar to the interaction between the query vector $Q$ and key vector $K$ in attention mechanisms, the input is weighted to affect the hidden state.

To accelerate training and alleviate the vanishing gradient problem, a residual connection is added between the SSM output and the convolution output:

$$Y_{residual} = Y + X_{conv\_activated} \tag{14}$$

Then, RMS normalization is applied for standardization to improve the model's stability, and a linear layer is used to map the output to the target dimension $D_{output}$.

$$Y_{normalized} = RMSNorm(Y_{residual}) \tag{15}$$

$$Y_{final} = Linear(Y_{normalized}) \tag{16}$$

**Dot-Product Attention:** Considering the imbalance in the importance of the rumor features output by the two directions of the Mamba2 Block, the network constructs a dot-product attention mechanism to weigh the outputs from both directions, enabling the model to assess the importance of the outputs from each direction. Finally, the weighted sum of the outputs from different directions is computed to enhance the model's ability to represent rumor information in the output features.

First, the dot product between the query $Q$ and the key $K$ is computed. Consider $x_{forward}$ as the query matrix $Q$ and $x_{backward}$ as the key matrix $K$. The dot product calculation is as follows:

$$scores = x_{forward}(x_{backward})^T \tag{17}$$

Then, the dot product result is scaled to avoid excessively large values. The scaling factor is the square root of the dimension of the key vector, $d_k$. After that, the scaled dot product is passed through the Softmax function to convert it into a probability distribution, $w_{forward}$, which represents the weight of $x_{forward}$ relative to $x_{backward}$. Similarly, $w_{backward}$ is also obtained:

$$scaled\_scores = \frac{scores}{\sqrt{d_k}} \tag{18}$$

$$w_{forward} = softmax(scaled\_scores)$$

Finally, the forward propagation $x_{forward}$ and backward propagation $x_{backward}$ are weighted and fused to obtain the final output $O$ of the network:

$$O = w_{forward} \cdot x_{forward} + w_{backward} \cdot x_{backward} \tag{19}$$

## 4.4. Rumor Locating and Marking

In the Rumor Locating and Marking module, we design a Skip-connection network that uses residual connections to maximize the retention and transmission of rumor-related information during the mapping from high-dimensional rumor features to low-dimensional label features. This skip connection effectively mitigate the issue of information loss in deep networks, ensuring the integrity of rumor features, and allowing precise retention of key semantics in the input data, even in complex dimensionality reduction tasks.

The network takes high-dimensional rumor features $O$ as input, and maps them to the first hidden layer through the first linear mapping layer, enhancing the feature expression capability with the non-linear activation function *SILU*, and preliminarily condensing the rumor features. The SILU activation function is advantageous for its smoothness, ability to prevent gradient explosion, effective handling of negative values, and capacity to accelerate model convergence, making it ideal for models requiring precise gradients and detailed representations.

$$x_1 = SILU(layer1(O)) \qquad (20)$$

To prevent potential information loss during the first linear mapping, the network employs a skip connection, concatenating the original input with the output of the first layer before feeding it into the second linear mapping, further refining comprehensive rumor features and capturing the complex semantic information in the input data.

$$x_2 = SILU\left(layer2(concat(O, x_1))\right) \qquad (21)$$

Finally, the output of the third layer is mapped to a low-dimensional label space through the output layer.

$$Emission\ Score = outputlayer(x_2) \qquad (22)$$

Direct dimensionality reduction of high-dimensional rumor features can lead to information loss during feature compression. Through the network's skip connections, these high-dimensional features are hierarchically compressed and transmitted, gradually mapped to a low-dimensional label space.

For the obtained low-dimensional label features (*Emission Score*), we employ Conditional Random Fields (CRF) to optimize the global matching of the output label sequence $Y = \{y_1, y_2, \dots, y_n\}$, maximizing the conditional probability P(Y|X). In this network, strong constraints in the labeling rules are incorporated by learning the transition matrix of parameters, and the Viterbi algorithm is employed to find the label sequence with the highest conditional probability. The conditional probability is defined as:

$$P(Y \mid X) = \frac{exp(Score(X,Y))}{\sum_{Y' \in y^n} exp(Score(X,Y'))} \qquad (23)$$

The denominator $Z(X) = \sum_{Y' \in y^n} exp(Score(X,Y'))$ is the normalization factor, ensuring that $P(Y|X)$ is a valid probability distribution. $X = \{x_1, x_2, \dots, x_n\}$ is the input feature sequence, and $Y = \{y_1, y_2, \dots, y_n\}$ is the label sequence.

The scoring function *Score(X, Y)* consists of two parts: the observation features (i.e., *Emission Score*) and the transition features.

$$Score(X,Y) = \sum_{i=1}^{n} Emission\ Score(x_i, y_i) + \sum_{i=1}^{n} Transfer\ Score(y_{i-1}, y_i) \qquad (24)$$

Here, $Emission\ Score(x_i, y_i)$ is the observation score between the input feature $x_i$ and the label $y_i$, reflecting the degree of alignment between the input and the label. $Transfer\ Score(y_{i-1}, y_i)$ is the transition score between adjacent labels $y_{i-1}$ and $y_i$, which represents the dependency between the labels. The transition score here corresponds to the corresponding value in the transition matrix obtained by CRF learning.

The Viterbi algorithm efficiently calculates the optimal label sequence using dynamic programming. First, the normalization factor $Z(X)$ is computed recursively to avoid enumerating all possible label combinations. Then, the Viterbi algorithm is used to find the label sequence that maximizes the conditional probability:

$$Y^* = \arg\max_{Y \in \mathcal{Y}^n} Score(X, Y) \tag{25}$$

In label feature classification, CRF considers the label dependencies of the entire sequence by using the transition relationships between labels and input features, maximizing the global probability of the true label sequence.

## 4.5. Loss Function

Log-Likelihood Loss, also known as Log Loss or Cross-Entropy Loss, is a commonly used loss function for classification problems. It measures the difference between the probability distribution output by the model and the true label's probability distribution. For a multi-class sequence problem, the log-likelihood loss can be expressed as:

$$\mathcal{L} = -\sum_{t=1}^{T} \log P(y_t \mid x_t; \theta) \tag{26}$$

Here, $T$ is the length of the sequence. $y_t$ is the true class of the t-th element in the sequence. $P(y_t \mid x_t; \theta)$ is the probability predicted by the model under parameter $\theta$ that $x_t$ belongs to class $y_t$. $\log$ represents the logarithm of the probability.

The objective of the proposed model is to maximize the conditional probability $P(Y|X)$ of the true label sequence $Y = \{y_1, y_2, \ldots, y_n\}$, which aims to maximize the probability of generating the true label sequence given the input sequence $X$. We optimize the conditional probability $P(Y|X)$ by minimizing the negative log-likelihood loss. Specifically, this involves maximizing the score of the true path while minimizing the sum of the scores of all possible paths.

$$\mathcal{L}_{\text{log-likelihood}} = -\sum_{(X,Y)} [Score(X, Y) - \log Z(X)] \tag{27}$$

Here, $Score(X|Y)$ is the score assigned by the model to the true label sequence $Y$ given the input sequence $X$, and $Z(X)$ is the normalization factor, also known as the partition function of the denominator graph, which computes the sum of scores of all possible label sequences to ensure the normalization of the probability distribution.

The model parameters are learned by minimizing the log-likelihood between the predicted label sequence and the true label sequence, with the Adam optimizer used to simultaneously update the parameters of all networks.

# 5. Experiments

## 5.1. A new dataset IR-WEIBO

In recent years, research on rumor detection has advanced. However, the majority of publicly available datasets mainly focus on determining whether a text is a rumor (i.e., a binary classification task), and lack support for detecting and labeling the specific rumor content within the text. This limitation makes it challenging for existing datasets to satisfy the demands of more detailed rumor analysis tasks.

To fill this gap, we have established a brand-new dataset called IR-WEIBO, which is dedicated to locating and marking specific rumor content in social media texts. This dataset is a dedicated resource for the specific task of detecting and labeling rumor content, which includes 3,200 text samples from the social media platform Weibo, derived from verified rumors. The labels in IR-WEIBO combine manual and automated labeling to ensure high quality and consistency. The IR-WEIBO dataset will be made available through an application-based access process.

The labels rules are as follows: "B-Rumor" marks the beginning of the rumor content, "I-Rumor" marks the rumor content, and "O" marks non-rumor content. These are represented by 0, 1, and 2 for "B-Rumor", "I-Rumor", and "O", respectively. The first column of the dataset contains the original text, and the second column contains the true labels. As shown in TABLE 1, the number of each label in the dataset.

TABLE 1 The statistics of the datasets.

| Statistic | B-Rumors | I-Rumors | O |
|---|---|---|---|
| IR-WEIBO | 3370 | 60409 | 247640 |

## 5.2. Implementation Details

We split the dataset into training, validation, and testing sets in a ratio of 8:1:1. The evaluation metrics include accuracy, precision, recall, and F1 score. We use the pre-trained BERT model "bert_base_chinese" as the word encoding tool. To better simulate real-world applications, the experiment does not remove the large number of non-rumor labels "O". In addition, in the Skip-connection network, the hidden layer dimensions for the layer are set to 512 and 256, respectively. Finally, during the model training, the Adam optimizer is used to optimize the model, with a learning rate set to 1e-5. All experiments are implemented based on PyTorch and Tesla V100-PCIE-32GB.

## 5.3. Baselines

Since there has been no prior research in this area, we made improvements to existing sequence labeling models in our experiment to enable them to perform rumor locating and marking tasks, and we demonstrate the effectiveness of our method through comparative experiments. The details are discussed below. Below, we briefly describe the seven methods that are being compared:

●The BERT+PLTE [Mengge *et al.*, 2020] for Rumor Locating and Marking: We modify the label mappings of the BERT+PLTE model to classify rumor-related spans and discard any unimplementable modules. By integrating the pre-trained BERT model for sequence labeling, we can leverage its rich contextual information to identify rumor-related entities, such as specific phrases or terms indicating rumors in social media posts. The PLTE network helps capture word boundary information, which aids in precise rumor span identification.

●The BERT+FLAT [Li *et al*., 2020] for Rumor Locating and Marking: For the rumor locating and marking task, we adjust the label mappings in the BERT+FLAT model to classify rumor-related spans. The FLAT method for position encoding is retained, as it enables the model to efficiently process lexicon-based cues that could be indicative of rumors while supporting parallel computation for faster inference. The BERT model integrates the contextual understanding needed to distinguish between rumor and non-rumor content in the IR-WEIBO dataset.

●DGLSTM-CRF [Jie and Lu, 2019] for Rumor Locating and Marking: The DGLSTM-CRF model is modified to focus on encoding the dependency relationships that highlight rumor-related elements in the text. We adapt the dependency-guided LSTM layers to emphasize features that are useful for identifying rumor spans, such as specific patterns of phrase dependencies that signal rumors. The CRF layer is modified to label the sequence with rumor/non-rumor tags instead of entity tags.

●The Star-GAT [Chen and Kong, 2021] for Rumor Locating and Marking: In adapting the Star-GAT model for rumor locating and marking, we modify the label mappings to focus on identifying rumor-related spans and discard tasks that are unrelated to the rumor detection. The model's graph attention network layer is used to capture the dependency relations between words that may be indicative of rumors, helping the model identify important spans. We treat rumor span identification as a binary classification task (rumor or non-rumor), with the attention mechanism assisting in focusing on the most relevant parts of the sentence.

●The WC-GCN [Tang *et al.*, 2020] for Rumor Locating and Marking: The WC-GCN model is adapted to focus on long-range dependencies relevant to rumor locating and marking. The global attention GCN block is fine-tuned to capture contextual information related to rumors across the entire text, enabling the model to learn effective node representations that highlight rumor-related entities or spans. The sequence labeling is modified to predict rumor-related boundaries instead of general entity labels.

●The RICON [Gu *et al.*, 2022] for Rumor Locating and Marking: The RICON model is adjusted for rumor locating and marking by modifying the regularity-aware and regularity-agnostic modules to detect spans related to rumors while avoiding an overemphasis on irrelevant span patterns. The model is designed to capture internal regularities of rumor-related spans and identify boundaries that correspond to rumor content, addressing the challenge of distinguishing rumors from non-rumors in the IR-WEIBO dataset.

## 5.4. Results and Discussion

We evaluate the performance of different methods on each metric for the labels "B-Rumor", "I-Rumor", and "O", with the results and comparisons presented in TABLE 2. Furthermore, we evaluate the accuracy of different methods in predicting the entire sentence sequence, with the results and comparisons presented in TABLE 3. To ensure fairness in the experiments, all comparison models are evaluated in the same experimental setting. In the experimental results, the bold data correspond to the best performance for each metric, and the horizontal line represents the second-best performance.

TABLE 2. Results of comparison among different models on IR-WEIBO datasets.

| Target \ Label \ Method | Accuracy | | | Precision | | | Recall | | | F1 Score | | |
|---|---|---|---|---|---|---|---|---|---|---|---|---|
| | B-R | I-R | O | B-R | I-R | O | B-R | I-R | O | B-R | I-R | O |
| BERT+PLTE | 0.625 | 0.749 | 0.832 | 0.663 | 0.687 | 0.732 | 0.620 | 0.703 | 0.823 | 0.641 | 0.695 | 0.775 |
| BERT+FLAT | 0.677 | 0.702 | 0.847 | 0.683 | 0.692 | 0.702 | 0.643 | 0.721 | 0.837 | 0.662 | 0.706 | 0.764 |
| DGLSTM-CRF | 0.747 | 0.826 | 0.863 | 0.762 | 0.779 | 0.893 | 0.698 | 0.774 | 0.866 | 0.729 | 0.776 | 0.879 |
| Star-GAT | 0.852 | 0.863 | 0.922 | 0.757 | 0.824 | 0.933 | 0.747 | 0.833 | 0.905 | 0.752 | 0.828 | 0.919 |
| WC-GCN | 0.824 | 0.853 | 0.902 | 0.832 | 0.877 | 0.958 | 0.799 | 0.852 | 0.911 | 0.815 | 0.864 | 0.934 |
| RICON | 0.832 | 0.869 | 0.894 | 0.802 | 0.874 | 0.968 | 0.797 | 0.863 | 0.925 | 0.800 | 0.868 | 0.946 |
| **Ours** | **0.893** | **0.905** | **0.983** | **0.885** | **0.898** | **0.970** | **0.859** | **0.886** | **0.974** | **0.872** | **0.892** | **0.972** |

TABLE 3. The accuracy of full-sentence locating and marking correctness for different models on the IR-WEIBO dataset.

| Method | BERT+PLTE | BERT+FLAT | DGLSTM-CRF | Star-GAT | WC-GCN | RICON | **Ours** |
|---|---|---|---|---|---|---|---|
| Accuracy | 0.453 | 0.528 | 0.622 | 0.698 | 0.666 | 0.679 | **0.738** |

The comparison of experimental results clearly demonstrates that the proposed model outperforms existing sequence labeling models across all metrics for each label. Specifically, for the "B-Rumor" label, which has a smaller sample size, the proposed model's accuracy, precision, recall, and F1 score are higher than those of other models by 0.041~0.268, 0.053~0.222, 0.06~0.239, and 0.072~0.231, respectively. The metrics for the "I-Rumor" and "O" labels are also higher than those of existing models by approximately 0.04~0.2 and 0.002~0.16, respectively. In addition, the proposed model significantly outperforms the compared sequence labeling models in terms of overall sentence labeling accuracy. The BERT+PLTE and BERT+FLAT models perform relatively poorly in accuracy, precision, and recall, especially with the low-sample labels "B-Rumor" and "I-Rumor", which may be due to their insufficient generalization ability in complex contexts. The DGLSTM-CRF model's mediocre performance on B-R and I-R tags may be attributed, on one hand, to the inadequate capacity of its tree-based encoding to handle contextual information, and on the other hand, to the potential loss of pivotal information during the transition from high-dimensional feature representations yielded by the DGLSTM to the low-dimensional label space. The RICON model's regularity-aware and regularity-agnostic modules may suffer from key information loss, which restricts the model's performance.

The proposed model outperforms existing sequence labeling models for the following reasons: The Mamba2 Block itself, by combining CNN and SSM, has the ability to handle long-range dependencies. The proposed model integrates this advantage and builds a bidirectional Mamba2 Block to further enhance its handling of long-range dependencies. Additionally, it improves the output's expressive capability through weighted summation with attention. Moreover, the Skip-connection Network designed in the model acquires low-dimensional label features, allowing the model to map high-dimensional rumor features to low-dimensional label features more completely.

## 5.5. Ablation Study

To validate the effectiveness of each module in Insight Rumors, we delete certain networks and key components to obtain simplified ablation variants of the model. The details of these simplified ablation variant models are described as follows:

IR-BERT deletes the BERT encoding part of the proposed model and directly uses Att_BiMamba2 for feature extraction of words in the text sequence. IR-Mamba2 replaces the Mamba2 Block in the proposed model with LSTM for rumor feature learning. IR-Dot-P-Att directly concatenates the bidirectional output results to obtain the final rumor features. IR-Skip-con deletes the Skip connection network designed in the proposed model and directly performs a single mapping to obtain label features. IR-CRF deletes the CRF and uses label features for a Max Pooling operation to obtain the labeling result.

TABLE 4. The accuracy of full-sentence locating and marking correctness for different models on the IR-WEIBO dataset.

| Target | Accuracy | | | Precission | | | Recall | | | F1 Score | | |
|---|---|---|---|---|---|---|---|---|---|---|---|---|
| Method \ Label | B-R | I-R | O | B-R | I-R | O | B-R | I-R | O | B-R | I-R | O |
| IR-BERT | 0.853 | 0.874 | 0.900 | 0.832 | 0.855 | 0.902 | 0.831 | 0.857 | 0.899 | 0.862 | 0.869 | 0.901 |
| IR-Mamba2 | 0.732 | 0.766 | 0.832 | 0.747 | 0.783 | 0.875 | 0.721 | 0.755 | 0.838 | 0.802 | 0.851 | 0.876 |
| IR-Dot-P-Att | 0.832 | 0.875 | 0.937 | 0.853 | 0.866 | 0.954 | 0.840 | 0.852 | 0.901 | 0.836 | 0.862 | 0.912 |
| IR-Skip-con | 0.888 | 0.893 | 0.940 | 0.883 | 0.892 | 0.952 | 0.849 | 0.876 | 0.974 | 0.871 | 0.8922 | 0.9653 |
| IR-CRF | 0.603 | 0.634 | 0.704 | 0.642 | 0.668 | 0.771 | 0.635 | 0.750 | 0.803 | 0.668 | 0.632 | 0.771 |
| **ALL** | **0.893** | **0.905** | **0.983** | **0.885** | **0.898** | **0.970** | **0.859** | **0.886** | **0.974** | **0.872** | **0.892** | **0.972** |

We compare these ablation variants with the complete Insight Rumors model under the same experimental conditions, as shown in TABLE 4 and TABLE 5. The table shows the labeling metrics for each label in the ablation variant models and the proposed model. The table shows the accuracy of complete sentence labeling for the text sequence in the ablation variant models and the proposed model.

TABLE 5. The accuracy of full-sentence locating and marking correctness for different ablation variants of Insight Rumors on the IR-WEIBO dataset.

| Method | IR-BERT | IR-Att_Mamba2 | IR-Dot-P-Att | IR-Skip-con | IR-CRF | **All** |
|---|---|---|---|---|---|---|
| Accuracy | 0.713 | 0.624 | 0.705 | 0.726 | 0.521 | **0.738** |

Upon analyzing the results in TABLE 4 and TABLE 5, the performance of all the ablation variants is inferior to that of the complete model. When the CRF network is added to the model, the evaluation results for each label show an improvement of approximately 0.1~0.3, indicating that adding strong constraint rules in the final labeling stage is crucial. When LSTM is used to replace Mamba2 as the rumor feature learning network, all metrics show a decrease of about 0.1~0.2, proving that the Mamba2 model's ability to learn rumor features is superior to that of the LSTM model. When the bidirectional Mamba2 model is equipped with dot-product attention to balance and fuse the outputs from both directions, the evaluation results for each label show significant improvement, and the accuracy of complete sentence evaluation is significantly increased. This suggests that the dot-product attention component can further enhance the bidirectional Mamba2 model's ability to express rumor information. After adding the Skip-connection network to the dimensionality reduction process from Att_BiMamba2 output to label features, the model's metrics for each label improve by at least 0.1. Although the improvement in sentence full evaluation accuracy is not significant, there is still a slight improvement, indicating that the Skip-connection network plays a role in ensuring the completeness of the mapping. Observing the ablation results without using the pre-trained model to obtain word vectors, we can clearly see a decline in the model's performance, indicating that the pre-trained BERT model still plays an important role in obtaining feature representations for each token.

# 6. Conclusions

In this paper, we tackle a critical gap in the field of rumor detection: the lack of in-depth detection and detailed locating and marking of specific rumor content. We propose the Insight Rumors model and create the first dataset IR-WEIBO for this research. The model performs in-depth detection and detailed locating and marking of specific rumor content by framing the task of detecting and labeling rumor content as a specialized sequence labeling problem. The proposed model combines a BERT encoder to encode all content in the text sequence, constructs a bidirectional Mamba2 network to learn high-dimensional rumor features, and employs dot-product attention with weighted summation to enhance the representation of rumor features. A skip-connection network is designed to map high-dimensional rumor features to low-dimensional label features, effectively ensuring the comprehensive mapping of rumor information. Finally, a Conditional Random Fields (CRF) is used to apply strong constraints, thereby improving the accuracy of the labeling. The Insight Rumors model effectively handles long-range dependencies, key information loss, and other issues in sequence labeling tasks, achieving, for the first time, effective detection and locating and marking of specific rumor content in text. Furthermore, experiments using the IR-Rumor dataset were conducted to evaluate the proposed model and compare its performance with several existing sequence labeling models. The results demonstrate that the proposed model outperforms existing models across all performance metrics for this task. Moreover, we conducted detailed ablation experiments on the proposed model to validate the effectiveness of each network and its components. In the future, we will continue optimizing this model and actively explore more effective approaches for this task.